\documentclass[12pt]{article} 
\usepackage{epsfig}
\usepackage{cite} 
\newcommand{\re}[1]{(\ref{#1})}
\newcommand{\ot}{\otimes}
\newcommand{\dq}{{{d\rule{15pt}{0pt}}\over {d\ln{Q^2}}}}
\newcommand{\msbar}{\overline{MS}}

\newcommand{\eq}{\begin{equation}} 
\newcommand{\eqx}{\end{equation}} 
\newcommand{\eqn}{\begin{eqnarray}} 
\newcommand{\eqnx}{\end{eqnarray}} 
\newcommand{\dt}{\Delta}

\newcommand{\sd}{\displaystyle} 
\newcommand{\nin}{\noindent} 
\newcommand{\as}{\widetilde{\alpha_s}(Q^2)} 

\begin{document} 
\nin 
%
%
\begin{center} 
{\Large \bf Double logarithms, $ln^2(1/x)$, and the NLO DGLAP evolution 
for the non-singlet component of the nucleon spin structure function, $g_1$.}\\ 
\vspace{10mm} 
 
{\Large Beata~Ziaja $^{\dag,\,\ddag,\,\ast}$} 
\footnote{e-mail: ziaja@tsl.uu.se}\\ 
 
{\footnotesize 

\vspace{3mm}  
           $^{\dag}$ \it Department of Theoretical Physics, 
	               Institute of Nuclear Physics,\\ 
                      \it Radzikowskiego 152, 31-342 Cracow, Poland\\ 
\vspace{3mm}
           $^{\ddag}$  \it Department of Biochemistry, Biomedical Centre,\\
                   \it Box 576, Uppsala University, S-75123 Uppsala, Sweden\\

\vspace{3mm} 
           $^{\ast}$ \it High Energy Physics, Uppsala University, 
	               P.O. Box 535, S-75121 Uppsala, Sweden\\ 
} 
\end{center} 
 
\vspace{5mm} 
\nin 
{\bf Abstract:} 
{\footnotesize  
Theoretical predictions show that at low values of Bjorken $x$ the spin 
structure function, $g_1$ is influenced by large logarithmic corrections, 
$ln^2(1/x)$, which may be predominant in this region. These corrections are also partially 
contained in the NLO part of the standard DGLAP evolution. Here we calculate 
the non-singlet component of the nucleon structure function, 
$g_1^{NS}=g_1^p-g_1^n$, and its first moment, using a unified evolution 
equation.
This equation incorporates the terms describing the NLO DGLAP evolution 
and the terms contributing to the $ln^2(1/x)$ resummation. 
In order to avoid double counting in the overlapping regions of the phase-space,
a unique way of including the NLO terms into the unified evolution equation is 
proposed. The scheme-independent results obtained from this unified evolution 
are compared to the NLO fit to experimental data, GRSV'2000. Analysis of the 
first moments of $g_1^{NS}$ shows that the unified evolution including 
the $ln^2(1/x)$ resummation goes beyond the NLO DGLAP analysis. Corrections 
generated by double logarithms at low $x$ influence the $Q^2$-dependence of the
first moments strongly.
} 
\vspace{6mm} 

{\bf 1. Introduction }

During the last years the interest in physics of polarized nucleon has increased
significantly. Experimental data describing the structure of the nucleon are now
available in various ranges of the momentum transfer, $Q^2$, and the 
Bjorken variable, $x$. New experiments emerge, and new data are expected to be available 
in the next future. The data from the region of low $x$ will be of special importance, 
since they may improve the estimation of the quark contribution to the total 
nucleon spin.
 
The previous data obtained by the EMC collaboration \cite{pemc} showed that the 
total participation of quarks in the proton spin was very small. This 
contradicted theoretical predictions, obtained from the Ellis-Jaffe sum rule. 
That sum rule expressed the moments of quark distributions in terms of the 
nucleon axial coupling constants. Following the Ellis-Jaffe sum rule, 
the quarks should participate in about one-fifth of the total spin of nucleon.
The discrepancy between the theoretical expectations and the experimental 
data has been often referred as the "proton spin crisis".
  
There have been several explanations proposed to clarify this. One of them 
refers to the not-yet-measured domain of very low values of Bjorken $x$, $x<10^{-3}$. 
Theoretical predictions show that at low $x$ the structure function of the 
polarized nucleon, $g_1(x,Q^2)$, is influenced by large logarithmic corrections,
$ln^2(1/x)$, \cite{BARTNS,BARTS}. 
As a consequence, large contributions to the moments of the structure functions
from this region are expected.

Analysis of the nucleon structure function at low $x$,
including the resummation of the logarithmic corrections, was
performed in \cite{kz,BARTNS,BARTS,blum95,blum9603,blum9606,blum99}. 
The logarithmic corrections were introduced for the unintegrated parton 
distributions in Refs.\ \cite{kz,BBJK,BZIAJA,MAN}. Those corrections
originated from the ladder and the non-ladder bremsstrahlung diagrams 
\cite{BARTNS,BARTS,QCD,QCD1}. 

The recursive equations for resummation of the logarithmic corrections were 
formulated. Afterwards, these evolution equations were completed with the 
DGLAP evolution terms calculated at the LO accuracy \cite{BBJK,kz}. Including 
the DGLAP evolution was neccessary for accurate description of the structure 
functions in the region of moderate and large values of $x$. 

Since the phase-space domains for the logarithmic
and the DGLAP contributions overlapped partially, subtractions from 
the evolution kernels were made in order to avoid double counting in those
regions. It was shown that including the $ln^2(1/x)$ terms into the evolution 
equations generated more singular behaviour of $g_1(x,Q^2)$ at low $x$, than 
that obtained after the standard LO DGLAP evolution. This led to the 
conclusion that the pure LO DGLAP evolution might be not complete at low x. 

On the other hand, the description of polarized data based exclusively
on the DGLAP evolution  has improved after including the DGLAP evolution terms 
calculated at the NLO accuracy \cite{reya,leader2}. 
These predictions described data accurately even in the region of low $x$ 
($x>10^{-4}$). However, the relation between
the NLO corrections and the $ln^2(1/x)$ resummation has not been clarified yet. 
It was expected that at low $x$ the NLO corrections contained in part the 
logarithmic terms but it was not clear to what extent those corrections 
overlapped.

Here we analyze the interplay between the NLO corrections and the 
corrections obtained from the resummation of double logarithms, $ln^2(1/x)$. 
We formulate unified evolution equations for the {\it unintegrated} parton 
distributions, where we include the NLO DGLAP terms.
Parton distributions are not physical observables and, in general, they 
depend on the factorization scheme applied to the NLO analysis (cf.\ 
\cite{blum95,blum9603,blum}). They yield (scheme-independent) physical
observables after a convolution with the Wilson coefficient functions (cf.\ 
Eq.\ \re{wilson}).   

In this study we will consider the non-singlet component of the nucleon 
structure function, $g_1^{NS}=g_1^p-g_1^n$. This component posseses a property
that it does not depend on the factorization scheme applied 
within the family of $\msbar$-like factorization schemes \cite{leader2,vann2}.
This occurs at the level of parton distributions already. 

We formulate the equations including the logarithmic corrections, $ln^2(1/x)$,
and the NLO corrections at the ${\overline {MS}}$ factorization scheme.
We introduce a unique way of performing subtractions in the evolution kernels,
so as to avoid double counting between the double logarithmic terms 
and the NLO ones in the overlapping regions of the phase-space. Predictions
for the non-singlet component of the nucleon spin structure function,
$g_1^{NS}(x,Q^2)$ and its first moment are then obtained. These results are 
compared with the NLO fit to the experimental data, GRSV'2000, \cite{grsv2000}.
Finally, our conclusions are listed.   

{\bf 2. Unintegrated parton distributions and DGLAP evolution}

Differential equations describing the evolution of standard (integrated) parton
distributions may be transformed into integral equations, if introducing
the unintegrated parton distributions. These distributions, $f_i(x,Q^2)$, 
are defined as:
\eq
\dt q_i(x,Q^2)= \dt q_i^{(0)}(x)+\int_{k_0^2}^{Q^2}\,{dk^2 \over k^2}\,
f_i(x,k^2),
\label{fi}
\eqx
where $\dt q_i(x,Q^2)$ denote the integrated parton distributions
with $\dt q_i^{(0)}(x)$ describing the contributions coming
from the non-perturbative region, $Q^2< k_0^2$. The cutoff $k_0^2$ is usually
$\sim 1$ GeV$^2$.

\noindent
The standard DGLAP equations written for the evolution of the integrated 
non-singlet parton distribution, $\dt q_{NS}$,  
\eqn
\dq \dt q_{NS}(x,Q^2)&=&\as\,(\dt P \ot \dt q_{NS})(x,Q^2),
\label{diff}
\eqnx
with $\sd \as\equiv{\alpha_s(Q^2) \over 2\pi}$, transform to the following 
integral equations:
\eqn
f_{NS}(x,Q^2)&=&\as(\dt P \ot \dt q_{NS}^{(0)})(x)
+\as\int_{k_0^2}^{Q^2} {dk^2 \over k^2}\,(\dt P \ot f_{NS})(x,k^2),
\label{int}
\eqnx
if the relation (\ref{fi}) is applied. In particular, we have: 
$\sd \dq \dt q_i(x,Q^2)=f_i(x,Q^2)$ at the LHS of 
Eq.\ (\ref{diff}), and, substituting \re{fi} to the RHS of Eq.\ \re{diff}, 
one obtains the RHS of Eqs.\ \re{int}.

Coefficients, $\dt P$, are the splitting functions calculated for the polarized 
deep inelastic scattering.
Symbol $\ot$ denotes the integral convolution of two functions, 
$(f \ot g)(x)=\int_0^1 dy \int_0^1 dz\,\delta(x-yz)\,f(y)\,g(z)$.
 
A complete description of the DGLAP evolution of polarized parton distributions
at the NLO accuracy was performed primarily in one of the $\msbar$ 
factorization scheme. The splitting functions describing the evolution of the 
non-singlet components and the singlet ones were firstly derived 
in \cite{vann1,vogel}, and they are listed in detail in the Appendix of Ref.\ 
\cite{reya}. 
The equations for the DGLAP evolution at the NLO order are structurally
identical with Eq.\ \re{diff}. The only difference is that the non-singlet
distribution, $\dt q_{NS}$, separates into two parts: the valence one, 
$\dt q_{NS}^+=\dt q -\dt {\bar q}$, and the asymmetric one, 
$\dt q_{NS}^-=\dt q + \dt {\bar q} - (\dt q^{\prime} + \dt {\bar q^{\prime}})$,
which have to be evolved with different kernels. Here we will evolve the 
non-singlet isospin-3 parton component,
\eq 
\dt q_{NS,3}=\dt u + \dt {\bar u} -\dt d - \dt {\bar d},
\label{qns3}
\eqx
using the asymmetric $\msbar$ evolution kernels \cite{reya}. This component 
contributes to $g_1^{NS}$.

NLO DGLAP evolution of parton distributions depends on the factorization scheme 
applied. In general, scheme dependence dissapears in physical observables, e.g. in 
the structure functions, $g_1(x,Q^2)$, which are combinations 
of the parton distributions convoluted with the Wilson coefficient functions, 
$\dt C_{q,g}(x)$: 
\eqn
g_1(x,Q^2)&=&{1 \over 2}\,\sum_{q=1}^{N_f}\,e^2_q
\left\{\left((\delta+\as\,\dt C_q)\ot (\dt q + \dt {\bar
q})\right)(x,Q^2)\right.\nonumber\\
&&\left.\rule{50pt}{0pt}+\as {1 \over N_f} (\dt C_g \ot \dt g)(x,Q^2)\right\},
\label{wilson}
\eqnx
where $\delta$ is the Dirac $\delta$-function, $\delta(x-1)$.
The coefficient functions calculated at the $\msbar$ scheme \cite{vann1} are 
taken from Ref. \cite{reya}. The constant $N_f$ denotes the number of active 
flavours. Here, $N_f=3$.

In case of the non-singlet parton distributions, the dependence on the factorization
scheme dissapears already at the level of the integrated parton
distributions, $\dt q_{NS}^{\pm}$, within the set of the $\msbar$-like schemes 
\cite{leader2,vann2}. The non-singlet component of the nucleon structure 
function is then, obviously, also scheme-independent, and it reads:
\eqn
g_1^{NS}(x,Q^2)&=&{1 \over 6}\,\left((\delta+\as\,\dt C_q)\ot 
\dt q_{NS,3}\right)(x,Q^2)
\label{wilsonns}
\eqnx
%
    

{\bf 3. Unified evolution equations including $ln^2(1/x)$ resummation.}

A detailed study of the nucleon structure function at low $x$ was  performed 
in \cite{kz}. There a complete resummation of the double logarithmic 
corrections, $\ln^2(1/x)$, was performed for the unintegrated parton 
distributions. 
Those corrections originated from summing up 
the ladder and the non-ladder bremsstrahlung diagrams \cite{BARTNS,BARTS} 
with the requirement of ordering the ratios, $k_n^2/x_n$, for exchanged partons. 
The ratio, $k_n^2/x_n$, corresponded to the Sudakov variable, $\beta$, 
the momentum, 
$k_n$, was the transverse momentum, and the fraction, $x_n$, was the 
longitudinal momentum fraction of the nth parton exchanged. The recursive
equations for the resummation of the logarithmic corrections were then 
formulated. The integrated parton distributions were obtained from the
unintegrated ones, using a relation similar to \re{fi}:
\eq
\dt q_i(x,Q^2)= \dt q_i^{(0)}(x)+\int_{k_0^2}^{W^2}\,{dk^2 \over k^2}\,
f_i(x^{\prime}=x(1+{k^2\over Q^2}),k^2),
\label{fix}
\eqx
where the phase-space was extended to $W^2=Q^2(1/x-1)$ corresponding to 
the total energy squared measured in the center-of-mass frame.

Following \cite{kz}, we combine the NLO DGLAP evolution and the logarithmic
resummation into a unified evolution equation, including both the
DGLAP kernels and the $ln^2(1/x)$ evolution kernels:
\eqn
f_{NS}(x,Q^2)&=&\as(\dt P \ot \dt q_{NS}^{(0)})(x)
+\as\int_{k_0^2}^{Q^2} {dk^2 \over k^2}\,(\dt P_{reg}\ot f_{NS})(x,k^2)\nonumber\\
&&\hspace*{10ex}{\bf(\hspace*{3ex}DGLAP\hspace*{3ex})}\nonumber\\
&+&
\as{4\over 3}
\int_{x^{\prime}}^1 {dz\over z}
\int_{Q^2}^{Q^2/z}
{dk^{2}\over k^{2}}
f_{NS}\left({x^{\prime}\over z},k^{2}\right)\nonumber\\
&&\hspace*{10ex}{\bf (\hspace*{3ex}Ladder\hspace*{3ex})}\nonumber\\
&-&\as
\int_{x^{\prime}}^1 {dz\over z}
\Biggl(
\Biggl[ \frac{\tilde  {\bf F}_8 }{\omega^2} \Biggr](z)
\frac{ {\bf G}_0 }{2\pi^2}
\Biggr)_{qq}
\int_{k_0^2}^{Q^2}
{dk^{2}\over k^{2}}
f_{NS}\left({x^{\prime}\over z},k^{2}\right)\nonumber\\
&-&\as
\int_{x^{\prime}}^1 {dz\over z}
\int_{Q^2}^{Q^2/z}
{dk^{2}\over k^{2}}
\Biggl(
\Biggl[\frac{\tilde   {\bf F}_8 }{\omega^2} \Biggr]
\Biggl(\frac{k^{2}}{Q^2}z \Biggr)\frac{ {\bf G}_0 }{2\pi^2}
\Biggr)_{qq}
f_{NS}\left({x^{\prime}\over z},k^{2}\right).\nonumber\\
&&\hspace*{10ex}{\bf(Non-ladder)}
\label{nloinf}
\eqnx
where $x^{\prime}=x(1+{k^2\over Q^2})$. For a detailed form of the kernels see
Appendix A. Matrices, ${\bf F}_8$ and ${\bf G}_0$, represent octet partial waves 
and colour factors respectively. They are described in detail in Appendix A. 
Symbol $\displaystyle \Biggl[{\widetilde { {\bf F}_8}/{\omega^2} }\Biggr](z)$ 
denotes the inverse Mellin transform of 
$\displaystyle {{\bf F}_8}/{\omega^2}$~:
\eq
\Biggl[{\widetilde { {\bf F}_8}/{\omega^2}  }\Biggr](z)=
\int_{\delta-i\infty}^{\delta+i\infty} {d\omega \over 2\pi i}
z^{-\omega}{{\bf F}_8(\omega)}/{\omega^2}.
\label{imellin}
\eqx	
with the integration contour located to the right of the singularities
of the function $\displaystyle {\bf F}_8(\omega)/{\omega^2}$.

Each of the kernels adds to the homogeneous part of the recursive evolution 
equation. In order to avoid double counting the NLO terms and the $ln^2(1/x)$ terms 
in the overlapping regions of the phase-space, we 
propose the following procedure of including the NLO terms into the 
evolution equations (cf. \cite{kz}). We divide the phase-space of these equations into two regions: 
(i) $k_0^2<k^2<Q^2$ and (ii) $Q^2<k^2<Q^2/z$. 
In the region (i) we keep all terms generated by the (non-ladder) double 
logarithmic corrections and add only the regular part of (the LO) and NLO 
DGLAP terms, $\dt P_{reg}$, i.e. this part which is not singular at 
$z\rightarrow 0$ hence does not generate any $ln^2(1/x)$ contributions. 
In the region (ii) both 
LO and NLO DGLAP terms do not appear, and we have there only contributions 
from the ladder and the non-ladder double logarithmic terms. This procedure is 
unique, and it uses a proven result of Refs.\ \cite{QCD,QCD1} that the 
$ln^2(1/x)$ resummation is complete after including the ladder and the 
non-ladder contributions.

While integrating the parton distributions over the extended phase-space, 
$k^2<W^2=Q^2(1/x-1)$ in \re{fix}, we may generate singular contributions
similar to those appearing in the Wilson coefficient, $\dt C_q$. 
Therefore, when calculating the non-singlet structure function, 
$g_1^{NS}$, we use only the regular part of the Wilson coefficient,
\eqn
\dt C_{q,reg}={4 \over 3}\left\{
(1+z^2)\left( {\ln{(1-z)} \over {(1-z)} } \right)_+
-{3 \over 2}\,{1 \over {(1-z)_+} }
+2 +z
-\left( {9 \over 2} + {\pi^2 \over 3} \right) \delta(1-z)
\right\}.
\label{rwilson}
\eqnx
A full form of the Wilson coefficient was taken from Ref.\ \cite{reya}.
Symbol $()_+$ is defined by the following convolution. 
A function $f(z)$ convoluted with a function $\left(g(z)\right)_+$ gives:
\eq
\int_0^1\,dz\,f(z)\,\left(g(z)\right)_+=
\int_0^1\,dz\,(f(z)-f(1))g(z)
\label{plus}
\eqx.      
     

{\bf 4. Results}

We solved numerically the evolution equation \re{nloinf} for 
the non-singlet isospin-3 parton component, $f_{NS,3}$. 

Two different parameterizations of the non-perturbative parton distribution,\\ 
$\dt q_{NS,3}^{(0)}(x)$, at $Q_0^2=1$ GeV$^2$ were used: 
(i) the GRSV'2000 fit \cite{grsv2000}, and (ii) a simple "flat" input
\eq
\Delta p_i^{(0)}(x)=N_i (1-x)^{\eta_i}
\label{dpi0}
\eqx
with $\eta_{u_v}=\eta_{d_v}=3$, $\eta_{\bar u} = \eta_{\bar s} = 7$
and $\eta_g=5$. Normalization constants $N_i$ were determined
by imposing the LO Bjorken sum rule for $\Delta u_v^{(0)}-\Delta d_v^{(0)}$,
and requiring that the first moments of all other distributions are the
same as those determined from  the QCD analysis \cite{STRATMAN}.
It was checked that parametrization (\ref{dpi0}) combined with
the unified equations gave reasonable description of the SMC data
on $g_1^{NS}(x,Q^2)$ \cite{BBJK} and on $g_1^p(x,Q^2)$ \cite{bkk}. 
This fit was also used in \cite{kz,ziaja} to investigate the magnitude of the 
double logarithmic corrections to the spin structure function 
of proton and to its first moment. 

The integrated parton distribution was then obtained by numerical integration
of $f_{NS,3}$:
\eq
\dt q_{NS,3}(x,Q^2)= \dt q_{NS,3}^{(0)}(x)+\int_{k_0^2}^{W^2}\,{dk^2 \over k^2}
\,f_{NS,3}(x^{\prime}=x(1+{k^2\over Q^2}),k^2),
\label{fix3}
\eqx
following Eq.\ \re{fix}.

\begin{figure}[t]
\begin{center}
\epsfig{width=12cm, file=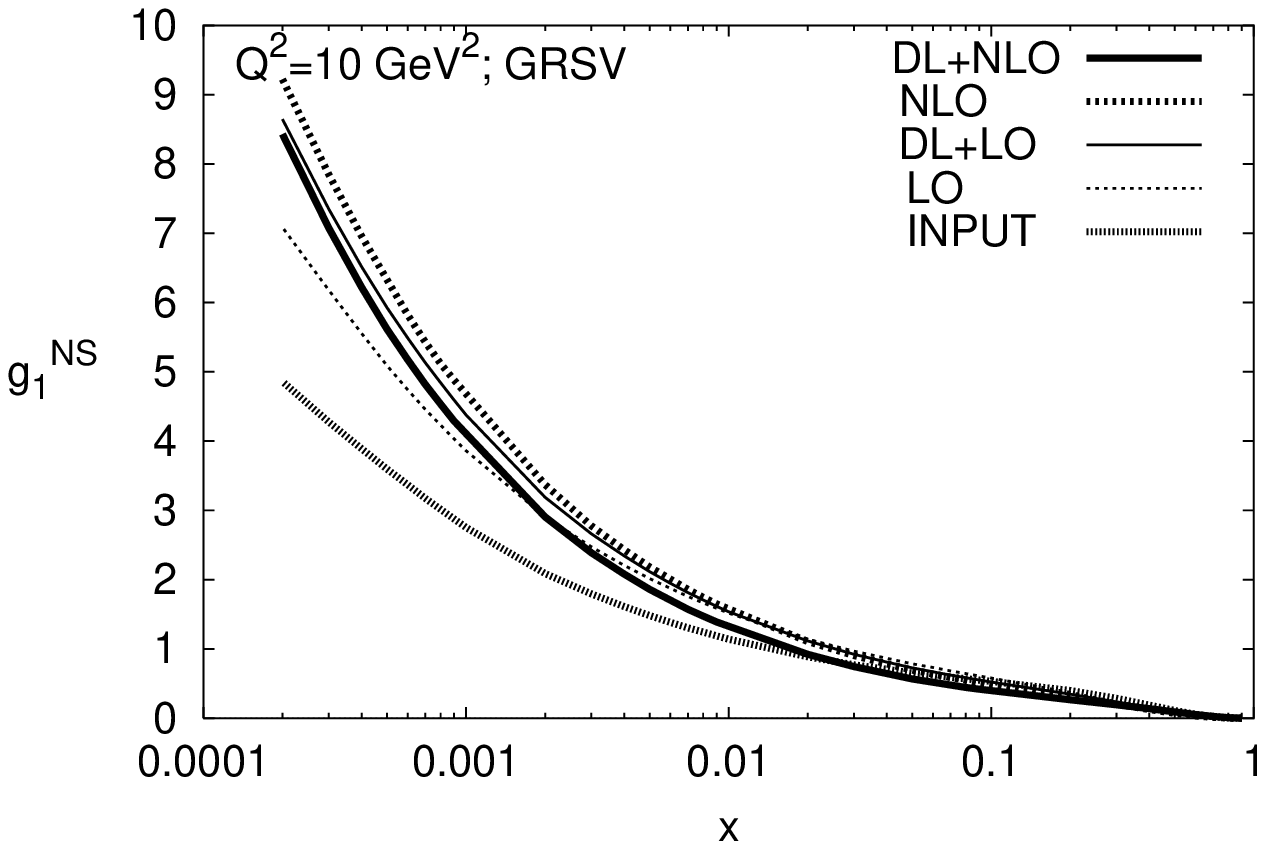}\\
\epsfig{width=12cm, file=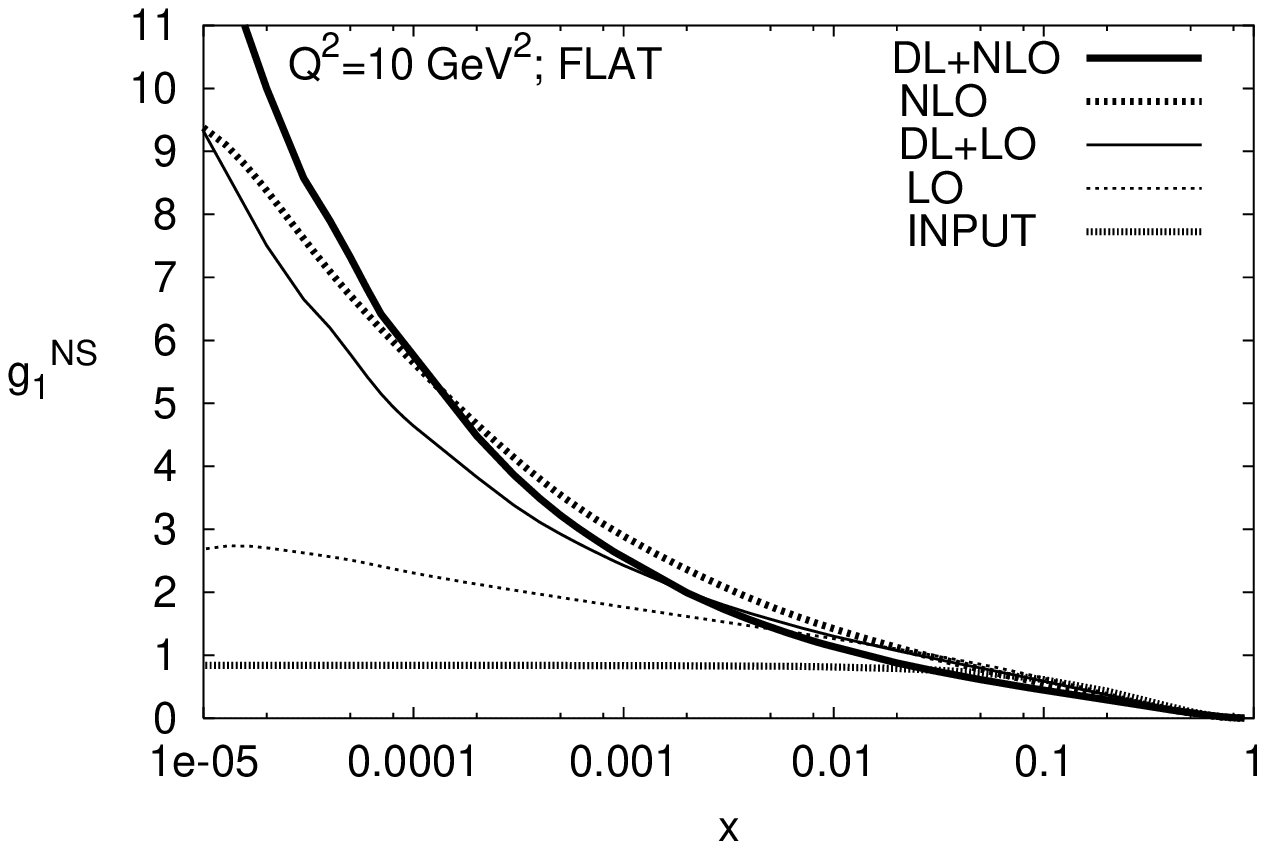}\\
\end{center}
\caption{Non-singlet component of the nucleon structure function, 
$g_1^{NS}=g_1^p-g_1^n$, plotted as a function of $x$ at the fixed 
$Q^2=10$ GeV$^2$.
Thick solid line ({\bf DL+NLO}) shows the results obtained by: (i) numerical 
solving Eq.\ \re{nloinf} for $f_{NS,3}$ with the input given (thick dotted line), 
(ii) numerical integration of $f_{NS,3}$ performed in order to obtain 
$\dt q_{NS,3}$, and (iii) numerical convolution of $\dt q_{NS,3}$ with 
the regular part of the Wilson coefficient function (Eq.\ \re{rwilsonns}).
Thick dashed line ({\bf NLO}) shows the NLO DGLAP evolution 
of the input. Thin lines correspond to the results: from the $ln^2(1/x)$ 
resummation including DGLAP terms at the LO accuracy (solid line, {\bf DL+LO}),
and from the pure LO DGLAP evolution (dashed line; {\bf LO}).
Upper plot shows the results from the GRSV'2000 input \cite{grsv2000}
($x>10^{-4}$), lower plot shows the results from the flat input \re{dpi0}
($x>10^{-5}$).
}
\label{fi1}
\end{figure}

Afterwards, we made a numerical convolution of $\dt q_{NS,3}$ with the
Wilson coefficient function, in order to obtain the non-singlet nucleon 
structure function, $g_1^{NS}$,
\eqn
g_1^{NS}(x,Q^2)&=&{1 \over 6}\,\left((\delta+\as\,\dt C_{q,reg})\ot 
\dt q_{NS,3}\right)(x,Q^2).
\label{rwilsonns}
\eqnx
Figs.\ \ref{fi1} a,b show the results. 

After evolving the GRSV input, the differences between curves resulting
from different approximations (DL+NLO, DL +LO, NLO, LO) were of the order of 
several percent at $x>10^{-3}$. The curve representing the unified DL+NLO 
evolution underestimated the pure NLO results. The same occured also for the evolution of 
the flat input \re{dpi0}, however, here the difference between the pure LO and 
other NLO results was described by the factor of $2$ at $x=10^{-4}$. 
In both cases the differences between the DL+NLO, DL+LO and NLO 
approximations manifested at very low x, $x\leq 10^{-4}$. 
This was clearly the region, where $ln^2(1/x)$ resummation dominated.
The results show that the shape of the input distribution influenced  
the evolution of the structure function strongly.
We checked that the DL+LO and LO results obtained for $g_1^{NS}$ 
are in a good agreement with the results of Ref.\ \cite{kz} 
(see Fig.\ 5 there).
 
Finally, we calculated the first moments of $g_1^{NS}$, $\Gamma_1^{NS}$,
starting from the inputs (i) and (ii). We estimated those moments in the 
following way. We
integrated the non-singlet structure function, $g_1^{NS}$, over $x$ in the 
region, $10^{-4}<x<1$. Since at $x<10^{-4}$ the function $g_1^{NS}$ was not 
known, we extrapolated $g_1^{NS}$ into this region with the fit, 
$x^{-\lambda}$, and then integrated it over $x$ at $0<x<10^{-4}$. It is worth
mentioning that the exponent, $\lambda$, was found to be $0.33-0.44$ at
$Q^2=2-10$ GeV$^2$ with different inputs. That was in agreement with 
the recent QCD-based estimate from Ref.\ \cite{knauf}, $\lambda=0.40\pm 0.29$.

\begin{figure}[t]
\begin{center}
\epsfig{width=12cm, file=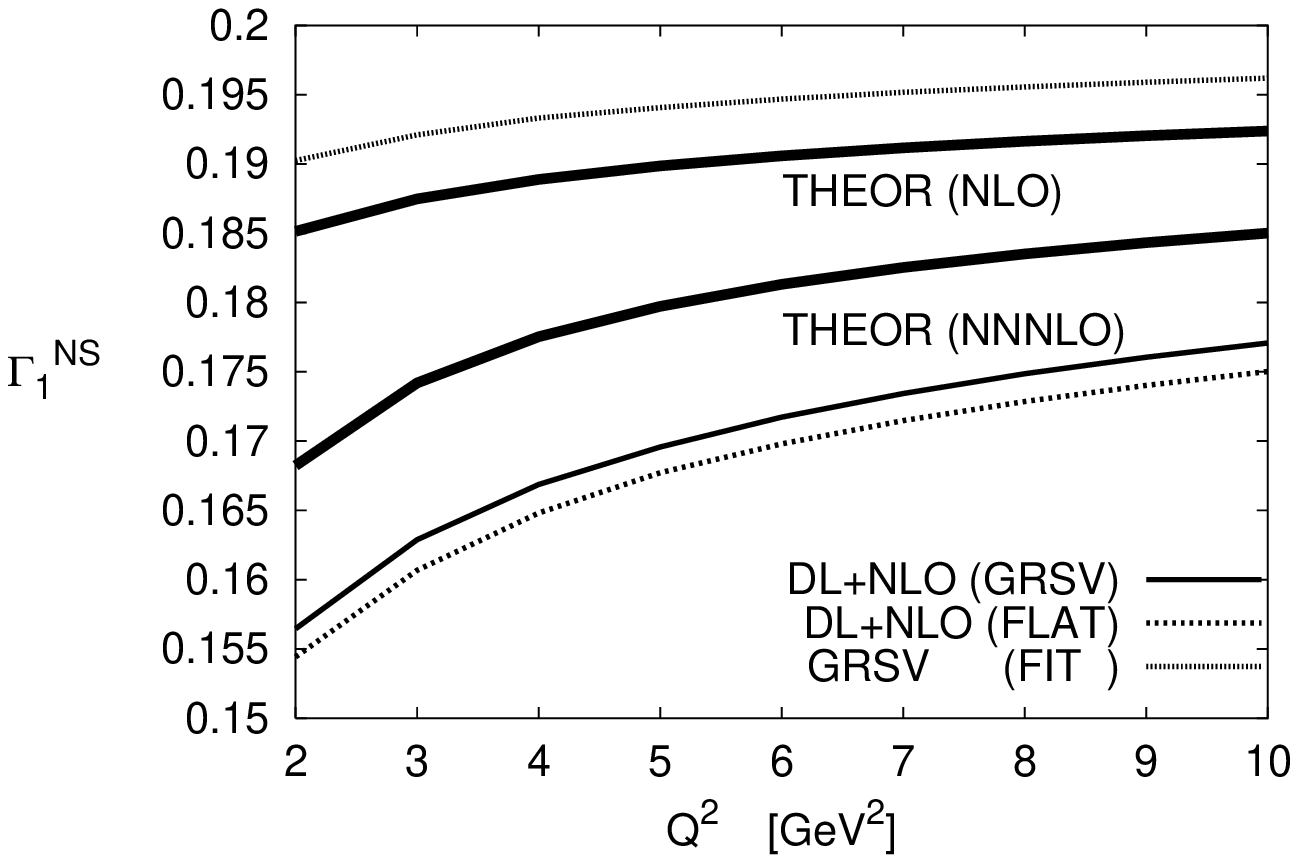}
\end{center}
\caption{First moments of the non-singlet component of the nucleon structure 
function, $g_1^{NS}=g_1^p-g_1^n$, as a function of $Q^2$ ($2<Q^2<10$ GeV$^2$).
Thin solid line shows the results from the GRSV non-perturbative input, dashed line
shows the results from the flat input \re{dpi0}. They are compared with the
theoretical predictions for the Bjorken sum rule (thick solid lines) calculated
at the NLO and NNNLO accuracy \cite{reya}, 
and with the first moment estimated from integrating out the 
NLO GRSV'2000 fit \cite{grsv2000} (dotted line).}
\label{fi2}
\end{figure}

The asymptotic corrections, added to the previous results, changed them at 
about one percent at most. The moments obtained are plotted in Fig.\ \ref{fi2}.  
Results obtained from the GRSV and the flat inputs, after
evolving them with the DL+NLO evolution equation, are similar.  They 
follow but underestimate the theoretical predictions
for $\Gamma_1^{NS}$ obtained from the QCD analysis at the NNNLO accuracy 
(Bjorken's sum rule \cite{reya}). This discrepancy is less then $8$ \%.
In contrast, $\Gamma_1^{NS}$ estimated from numerical integration of the 
GRSV'2000 fit \cite{grsv2000} follows Bjorken's sum rule calculated at the
NLO accuracy, $O(\alpha_s)$. This clearly suggests that the unified evolution,
DL+NLO, goes beyond the standard NLO DGLAP analysis, and that the corrections 
generated by double logarithms in the region of low $x$ influence 
the behaviour of the first moments significantly .   
%
%
\noindent
\noindent

{\bf 5. Conclusions}

We calculated the non-singlet component of the nucleon structure function,
$g_1^{NS}=g_1^p-g_1^n$, and its first moment, using a unified evolution equation.
This equation incorporated both the terms describing the NLO DGLAP evolution 
and the terms which contributed to the $ln^2(1/x)$ resummation. A unique way of
including the NLO terms into the unified evolution equation was proposed, 
in order to avoid double counting in the overlapping regions of the phase-space.
The results obtained from different approximations were 
similar, and the minor differences manifested at very low $x$,
$x\leq 10^{-4}$, where the $ln^2(1/x)$ resummation dominated. This effect 
depended on the shape of the non-perturbative input distribution, 
$\dt q_{NS}^{(0)}(x)$. First moments obtained from different non-perturbative 
inputs evolved with DL+NLO evolution equations were similar. These moments
followed Bjorken's sum rule calculated at the NNNLO accuracy. The discrepancy 
was at about $8$ \% at most. This shows that the unified 
evolution goes beyond the NLO DGLAP analysis. The $ln^2(1/x)$
corrections influence the $Q^2$-dependence of the first moments strongly.

Our results are scheme-independent within the $\msbar$-like family of
factorization schemes \cite{leader2,vann2}.

 
\section*{Acknowledgments} 
 
I am grateful to J. Kwieci\'nski for inspiring discussions and illuminating 
comments.
This research has been supported in part by the Polish Committee for Scientific       
Research with grants 2 P03B 05119, 2PO3B 14420 and European 
Community grant 'Training and Mobility of Researchers', Network 'Quantum      
Chromodynamics and the Deep Structure of Elementary Particles'      
FMRX-CT98-0194. B.\ Z.\ was supported by the Wenner-Gren Foundations.
%

\section*{Appendix A}

Here a brief description of the evolution kernels of Eq.\ \re{nloinf} 
is given. DGLAP kernels were taken from Ref.\ \cite{reya}. The full DGLAP kernel 
$\dt P$ includes both the LO and NLO terms:
\eq
\dt P=\dt P_{LO} + \as\, \dt P_{NLO}.
\label{dglap}
\eqx
In the homogeneous term, $\displaystyle \as\int_{k_0^2}^{Q^2} {dk^2 \over k^2}
\,(\dt P_{reg}\ot f_{NS})(x,k^2)$, appearing in \re{nloinf}, only the regular 
part of the full kernel
is included. This is to avoid double counting the NLO terms and $\ln^2(1/x)$
terms in the region of the phase-space, $k_0^2<k^2<Q^2$.

Ladder kernels corresponding to the LO DGLAP kernels taken at $z=0$ \cite{kz} 
generate double logarithmic corrections in the region of $Q^2<k^2<Q^2/z$. 

Non-ladder kernels were obtained in Ref.\ \cite{kz} from the infrared evolution
equations for the singlet partial waves ${\bf F_0}$, $\bf F_8$ 
\cite{BARTNS,BARTS,QCD,QCD1}. In \cite{kz} we noticed that extending
the kernel of the double logarithmic evolution equations from the ladder one, 
\eq
\as \dt P_{qq}/\omega\nonumber,
\eqx
to the modified one,
\eq
\as \left( \dt P_{qq}/\omega
-({\bf F_8}(\omega)\,{\bf G_0})_{qq}/(2\pi^2\omega^2) \right)\nonumber, 
\eqx
gave a proper anomalous dimension as derived from the infrared 
evolution equations.

Matrix ${\bf G}_0$ contained colour factors resulting from attaching the soft 
gluon to external legs of the scattering amplitude~:
\eqn
{\bf G}_0 &=&\left( \begin{array}{cc}  {N^2-1 \over 2N} & 0  \\
                                                    0 & N  \\ \end{array} 
						    \right ),
\label{g0}
\eqnx
where $N$ was the number of colours.

Further, it was checked that the Born approximation of ${\bf F_8}$,
\eq
{\bf F}_8^{Born}(\omega)\approx 8\pi^2 \as \frac{{\bf M}_8}{\omega}.
\eqx
gave accurate results for the DL evolution.
Martix ${\bf M}_8$ was a splitting function matrix in colour octet $t-$channel, 
\eqn
{\bf M}_8 &=&\left( \begin{array}{cc} -{1 \over 2N} & -{N_F \over 2}\\
                                                N & 2N \\ \end{array} \right ).
\eqnx
The inverse Melin transform of ${\bf F}_8^{Born}(\omega)$ then read~:
\eq
\Biggl[\frac{\tilde {\bf F}_8^{Born}}{\omega^2}\Biggr](z)=
4\pi^2 \as {\bf M}_8 ln^2 (z).
\label{born}
\eqx

The evolution equation \re{nloinf} includes the non-ladder corrections in 
the Born approximation \re{born}.    

%

\begin{thebibliography}{10}

\bibitem{pemc}
J.~Ashman {et al.}
\newblock {EMC}.
\newblock {\em Phys. Lett. B}, 206:364, 1988.

\bibitem{BARTNS}
J.~Bartels, B.~I. Ermolaev, and M.~G. Ryskin.
\newblock {\em Z. Phys. C}, 70:273, 1996.

\bibitem{BARTS}
J.~Bartels, B.~I. Ermolaev, and M.~G. Ryskin.
\newblock {\em Z. Phys. C}, 72:627, 1996.

\bibitem{kz}
J.~Kwieci\'nski and B.~Ziaja.
\newblock {\em Phys. Rev. D}, 60:054004, 1999.

\bibitem{blum95}
J.~Bl\"umlein and A.~Vogt.
\newblock {\em Phys. Lett. B}, 370:149, 1996.

\bibitem{blum9603}
J.~Bl\"umlein and A.~Vogt.
\newblock {\em Acta Phys. Pol. B}, 27:1309, 1996.

\bibitem{blum9606}
J.~Bl\"umlein and A.~Vogt.
\newblock {\em Phys. Lett. B}, 386:350, 1996.

\bibitem{blum99}
J.~Bl\"umlein.
\newblock {Lectures given at Ringberg Workshop: New Trends in HERA Physics
  1999, Ringberg Castle, Tegernsee, Germany, 30 May - 4 Jun 1999. In *Tegernsee
  1999, New trends in HERA physics* 42-57. Eds. G. Grindhammer, B. Kniehl and
  G. Kramer (Springer, Berlin 1999).}
\newblock {\em {hep-ph/9909449}}.

\bibitem{BBJK}
B.~Bade\l{}ek and J.~Kwieci\'nski.
\newblock {\em Phys. Lett. B}, 418:229, 1998.

\bibitem{BZIAJA}
J.~Kwieci\'nski and B.~Ziaja.
\newblock {\em {Proceedings of the Workshop "Physics with polarized protons at
  HERA"; DESY March-September 1997 (editors A. De Roeck, T. Gehrmann);DESY
  Proceedings 1998 }}, 1997.

\bibitem{MAN}
S.~I. Manayenkov and M.~G. Ryskin.
\newblock {\em {Proceedings of the Workshop "Physics with polarized protons at
  HERA"; DESY March-September 1997 (editors A. De Roeck, T. Gehrmann);DESY
  Proceedings 1998 }}, 1997.

\bibitem{QCD}
R.~Kirschner and L.~N. Lipatov.
\newblock {\em Nucl. Phys. B}, 213:122, 1983.

\bibitem{QCD1}
R.~Kirschner.
\newblock {\em Z. Phys. C}, 67:459, 1995.

\bibitem{reya}
B.~Lampe and E.~Reya.
\newblock {\em Phys.Rept.}, 332:1, 2000.

\bibitem{leader2}
E.~Leader, A.~V. Sidorov, and D.~B. Stamenov.
\newblock {\em Phys. Lett. B}, 445:232, 1998.

\bibitem{blum}
J.~Bluemlein, V.~Ravindran, and W.~L. van Neerven.
\newblock {\em Nucl. Phys. B}, 586:349, 2000.

\bibitem{vann2}
E.~B. Zijlstra and W.~L. van Neerven.
\newblock {\em Nucl. Phys. B}, 417:61, 1994.

\bibitem{grsv2000}
M.~Glueck, E.~Reya, M.~Stratmann, and W.~Vogelsang.
\newblock {\em Phys.Rev.D}, 63:094005, 2001.

\bibitem{vann1}
R.~Mertig and W.~L. van Neerven.
\newblock {\em Z. Phys. C}, 70:637, 1996.

\bibitem{vogel}
W.~Vogelsang.
\newblock {\em Phys. Rev. D}, 54:2023, 1996.

\bibitem{STRATMAN}
M.~Stratmann.
\newblock {\em {hep-ph/9710379}}, 1997.

\bibitem{bkk}
B.~Bade\l{}ek, J.~Kiryluk, and J.~Kwieci\'nski.
\newblock {\em Phys.Rev.D}, 61:014009, 2000.

\bibitem{ziaja}
B.~Ziaja.
\newblock {\em Acta Phys. Polon. B}, 32:2863, 2001.

\bibitem{knauf}
A.~Knauf, M.~Meyer-Hermann, and G.~Soff.
\newblock {\em hep-ph/0206204}.

\end{thebibliography}

\end{document}